# Do we agree on user interface aesthetics of Android apps?

Christiane G. von Wangenheim*[a], João V. Araujo Porto[a], Jean C.R. Hauck[a],
Adriano F. Borgatto[a]
[a]Department of Informatics and Statistics
Federal University of Santa Catarina (UFSC)
Florianópolis/Brazil
c.wangenheim@ufsc.br, joao.porto@grad.ufsc.br, jean.hauck@ufsc.br, adriano.borgatto@ufsc.br

**Abstract**
**Context:** Visual aesthetics is increasingly seen as an essential factor in perceived usability, interaction, and overall appraisal of user interfaces especially with respect to mobile applications. Yet, a question that remains is how to assess and to which extend users agree on visual aesthetics.
**Objective:** This paper analyzes the inter-rater agreement on visual aesthetics of user interfaces of Android apps as a basis for guidelines and evaluation models.
**Method:** We systematically collected ratings on the visual aesthetics of 100 user interfaces of Android apps from 10 participants and analyzed the frequency distribution, reliability and influencing design aspects.
**Results:** In general, user interfaces of Android apps are perceived more ugly than beautiful. Yet, raters only moderately agree on the visual aesthetics. Disagreements seem to be related to subtle differences with respect to layout, shapes, colors, typography, and background images.
**Conclusion:** Visual aesthetics is a key factor for the success of apps. However, the considerable disagreement of raters on the perceived visual aesthetics indicates the need for a better understanding of this software quality with respect to mobile apps.

## 1. Introduction

An integral part of usability as a software product quality is the aesthetics of user interfaces [1]. Visual aesthetics refers to the beauty or the pleasing appearance of user interfaces of an interactive software system [1,2]. It is considered a key factor on the perceived and effective usability of the user interface, subjective user satisfaction, trust & credibility, and preference [3]. Although, there exist research investigating first impressions on the aesthetics of desktop systems or web sites and how it affects user experience and usability [4], relatively little research has addressed mobile applications [5]. Yet, with well over millions of apps in the major stores, first impressions of visual aesthetics become decisive in the choice of apps. Therefore, user interfaces of mobile apps must be designed to highlight the most commonly used functions and to make most effective use of the screen and the mobile paradigm, including user input and associated sensor information and media access [6].

Yet, how to assess the visual aesthetics of apps remains a question during the user interface design process [3]. Human judgment, especially when it comes to evaluating aesthetics, remains by nature an intrinsically subjective process [3]. It seems, that to some extent, we can indeed generalize what users consider usable and attractive defining universally accepted norms. Some aspects, however, are certainly a matter of personal taste or strongly influenced by gender or cultural values. However, for further research, we need to know to which degree the perception of visual aesthetics of mobile app user interfaces is consistent and not being an idiosyncratic result of a user's subjective judgment.

## 2. Definition and execution of the study

Therefore, we conduct a study to analyze the inter-rater agreement on visual aesthetics of user interfaces, focusing on Android apps, being currently the most prominent mobile platform. Inter-rater agreement is defined as the degree to which the subjective ratings on visual aesthetics are identical.

We performed a survey with 10 participants recruited from the Software Quality Group and the Hiperlab at the Federal University of Santa Catarina[1]. For the study, we randomly selected 100 different user interfaces:
- 50 images of user interfaces (UIs) of Android apps from the Rico dataset [7], one of the largest repositories containing 72k unique UI screens of Android apps from the Google Play Store.
- 50 images of user interfaces (UIs) of apps from the App Inventor gallery, being one of the most popular programming environments for teaching app development worldwide.

Using the selected images, we automatically created an online form, including an explanation of the study, consent form and 100 randomly ordered questions, each presenting:
- the image of a user interface (239x425 pixels with a resolution of 96dpi in 24-bit true color), and
- a single-item question on the perceived visual aesthetics with the nominal response options beautiful/ugly.

Participants responded the online questionnaire remotely on their own computer. Although, no time constraint was imposed to participants, in general, participants spent about 15 minutes responding the questionnaire.

## 3. Results

### 3.1. How consistent are perceptions on user interface aesthetics of Android apps?

Analyzing the total sum of the ratings, we observe that participants rated the UIs almost two times more as ugly (645 ratings) than beautiful (355 ratings). This seems to be a general tendency on this dataset, as all participants rated more interfaces as ugly (Figure 1). However, some participants demonstrated more balanced ratings than others, who rated almost all UIs as ugly.

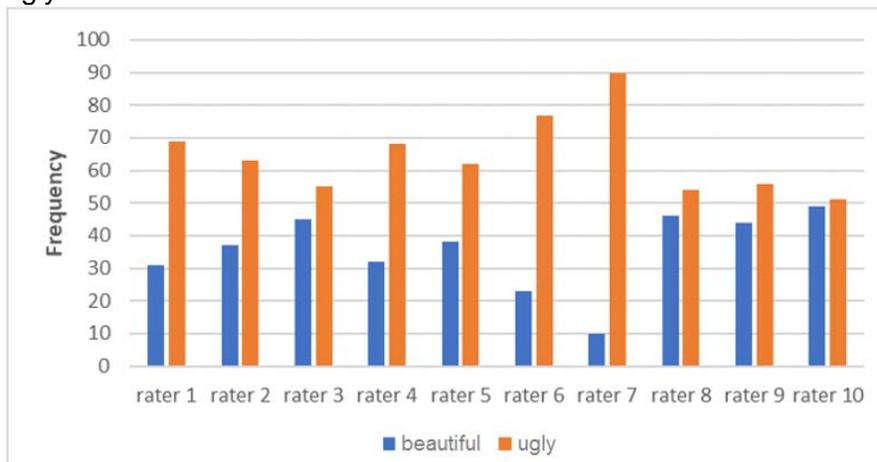

Figure 1. Frequency distribution per rater

---

[1] This study has been approved by the Ethics Committee of the Federal University of Santa Catarina (no. 2.903.849).

As we are interested in exploring design aspects and not the raters and their use of the scale, we analyze the degree of agreement with respect to the user interfaces (Figure 2). Yet, only with respect to 19 interfaces all raters agreed on aesthetics (either beautiful or ugly). Raters disagreed on the aesthetics of a considerable number of interfaces, reaching the extreme disagreement of 50% perceiving the interfaces as beautiful and 50% as ugly with regard to 9 UIs.

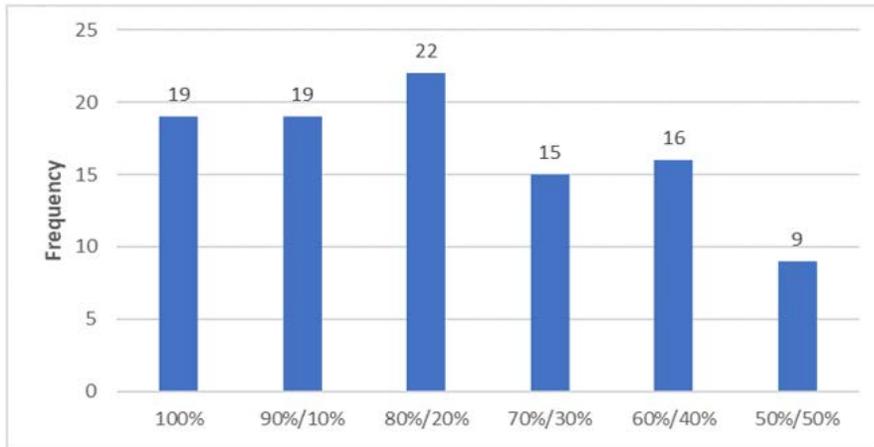

Figure 2. Frequency of degree of agreement

In order to analyze the reliability of agreement between 10 raters assigning categorical ratings, we used Fleiss' kappa. The resulting Fleiss' kappa = 0.302 indicates only a fair agreement between the raters following the interpretation suggested by Landis & Koch [8]. Analyzing in detail the inter-rater agreement reliability by calculating Cohen's kappa for each pair of two raters (Figure 3) shows that the inter-rater agreement reliability is also very low between most pairs of raters with few exceptions demonstrating moderate levels of agreement.

|         | Rater1 | Rater2 | Rater3 | Rater4 | Rater5 | Rater6 | Rater7 | Rater8 | Rater9 | Rater10 |
|---------|--------|--------|--------|--------|--------|--------|--------|--------|--------|---------|
| Rater1  | 1.00   | 0.29   | 0.60   | 0.53   | 0.51   | 0.26   | 0.34   | 0.20   | 0.26   | 0.27    |
| Rater2  | 0.29   | 1.00   | 0.19   | 0.25   | 0.23   | 0.27   | 0.27   | 0.04   | 0.07   | 0.12    |
| Rater3  | 0.60   | 0.19   | 1.00   | 0.64   | 0.81   | 0.49   | 0.25   | 0.31   | 0.43   | 0.26    |
| Rater4  | 0.53   | 0.24   | 0.64   | 1.00   | 0.64   | 0.62   | 0.40   | 0.11   | 0.35   | 0.19    |
| Rater5  | 0.51   | 0.22   | 0.81   | 0.64   | 1.00   | 0.56   | 0.32   | 0.24   | 0.44   | 0.24    |
| Rater6  | 0.26   | 0.27   | 0.49   | 0.62   | 0.56   | 1.00   | 0.49   | 0.16   | 0.22   | 0.09    |
| Rater7  | 0.34   | 0.26   | 0.25   | 0.40   | 0.32   | 0.49   | 1.00   | 0.10   | 0.15   | 0.13    |
| Rater8  | 0.20   | 0.04   | 0.31   | 0.11   | 0.24   | 0.16   | 0.10   | 1.00   | 0.11   | 0.34    |
| Rater9  | 0.27   | 0.07   | 0.43   | 0.35   | 0.44   | 0.23   | 0.16   | 0.11   | 1.00   | 0.22    |
| Rater10 | 0.27   | 0.11   | 0.26   | 0.19   | 0.24   | 0.09   | 0.13   | 0.34   | 0.21   | 1.00    |

Figure 3. Cohen's kappa analyzing the inter-rater agreement reliability between pairs of raters

### 3.2 Which design aspects seem to influence the consistency of perceptions?

Comparing interfaces that have been perceived as beautiful/ugly by all participants, we can observe several differences with respect to some layout aspects, including balance, equilibrium, symmetry, simplicity (Figure 4). There are also significant differences with respect to the color palettes, considering the kind of colors as well as the quantity of different colors used. The choice of fonts and shapes as well as their variety also seems to be relevant. Furthermore, the usage of background images, especially when reducing legibility, seems to influence negatively the perception of visual aesthetics.

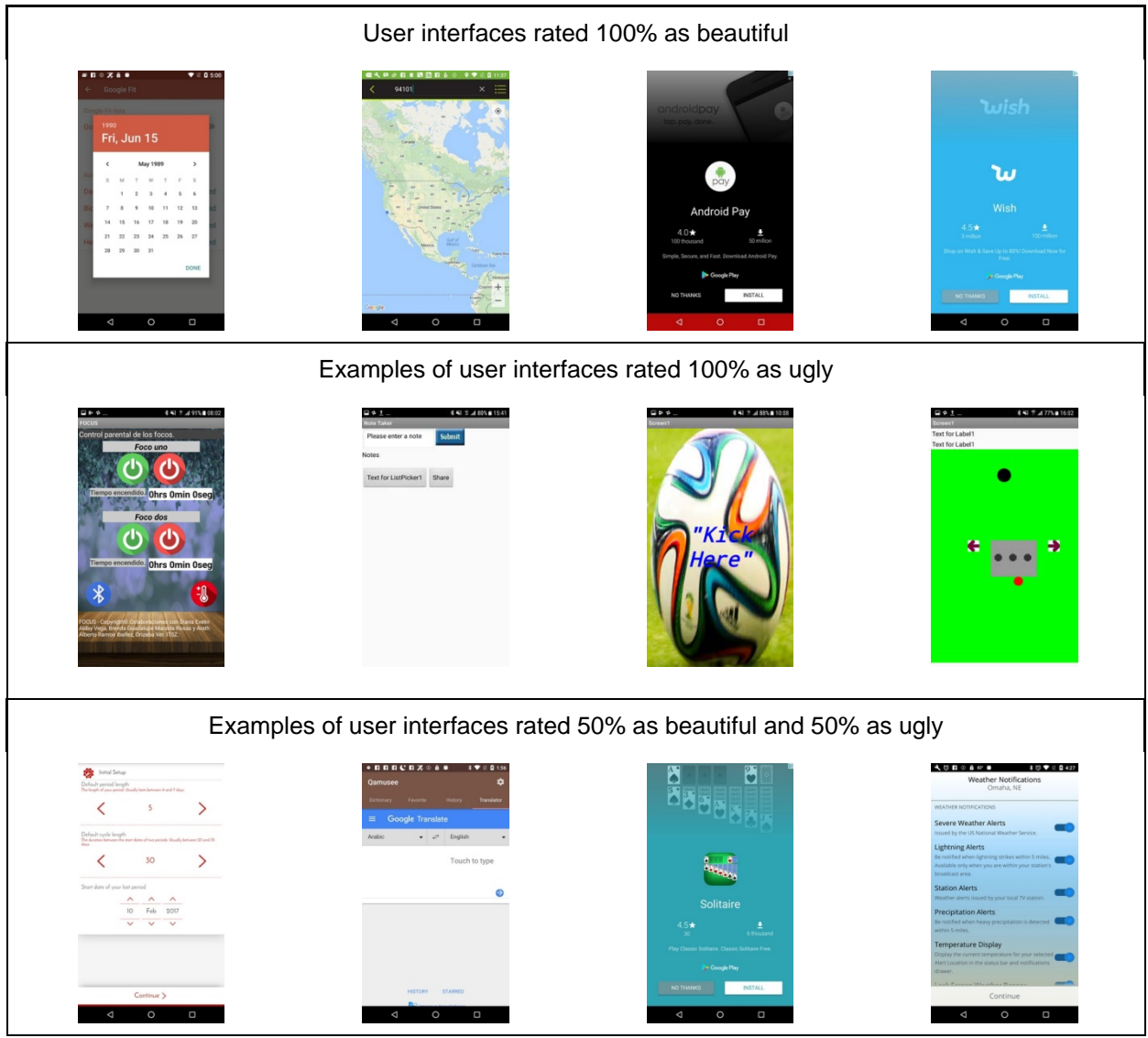

Figure 4. Examples of interfaces with different degrees of agreement

Analyzing interfaces with divergent perceptions, these differences seem to be related to the same aspects yet in a more subtle way. Thus, it seems that rather than one specific aspect causing disagreement, the perception seems to vary among raters, some, e.g., considering interfaces still as beautiful, even if some of these visual aspects are represented imperfectly (e.g., using non-combining colors or unbalanced designs).

**Limitations of the study.** Being an explorative research, this study has certain limitations. We used a single-item measure on a nominal scale assuming the measured construct being one-dimensional, as there still does not exist a consensus on the dimensionality nor a scale and wording. In terms of statistical significance, we obtained a small sample size, yet, robust to estimate inter-rater agreement. The relative homogeneous nature of the participants restricts the generalizability of the results. However, as even in this limited context, we observed a considerable disagreement, it is only expected to increase with participants from more heterogeneous backgrounds. To minimize threats related to data analysis, we performed a statistical evaluation using techniques for assessing disagreement between respondents for data on a nominal scale and carefully analyzed the data to evaluate the inter-rater agreement from different angles.

## 5. Conclusions

Our findings suggest that, although, there exist a certain degree of agreement, raters also disagree on the visual aesthetics on a considerable number of UIs of Android apps. We found that these differences are related to subtle differences with respect to diverse visual aspects, such as layout, shapes, color, typography and background images. In contrast to other studies with respect to web sites that report at least a medium degree of agreement [3], the results of our research indicate the need for further research in the area of mobile apps pointing out different patterns of perceptions. This understanding is essential in order to develop rationales and guidelines on visual aesthetics of UI of Android apps, as well as the identification of aesthetic dimensions and factors.

**Acknowledgements**

This work was supported by the CNPq (*Conselho Nacional de Desenvolvimento Científico e Tecnológico*), an entity of the Brazilian government focused on scientific and technological development.